\documentclass[prl,twocolumn]{revtex4}
\usepackage{amsmath,amssymb,mathrsfs,braket,hyperref,paralist,bm,graphicx,xcolor,soul,siunitx}

\begin{document}
\title{Observation of Quantum Criticality and Tomonaga-Luttinger Liquid \\ in One-dimensional Bose Gases}
\author{Bing Yang$^{1,3,*}$}
\author{Yang-Yang Chen$^{2,*}$}
\author{Yong-Guang Zheng$^{1,3}$}
\author{Hui Sun$^{1,3}$}
\author{Han-Ning Dai$^{1,3}$}
\author{Xi-Wen Guan$^{2,4,\dagger}$}
\author{Zhen-Sheng Yuan$^{1,3,5,6,\dagger}$}
\author{Jian-Wei Pan$^{1,3,5,6,\dagger}$}
\affiliation{$^{1}$Hefei National Laboratory for Physical Sciences at Microscale and Department of Modern Physics, University of Science and Technology of China, Hefei, Anhui 230026, China}
\affiliation{$^2$State Key Laboratory of Magnetic Resonance and Atomic and Molecular Physics, Wuhan Institute of Physics and Mathematics, Chinese Academy of Sciences, Wuhan 430071, China}
\affiliation{$^{3}$Physikalisches Institut, Ruprecht-Karls-Universit\"{a}t Heidelberg, Im Neuenheimer Feld 226, 69120 Heidelberg, Germany}
\affiliation{$^4$Department of Theoretical Physics, Research School of Physics and Engineering,  Australian National University, Canberra ACT 0200, Australia}
\affiliation{$^5$CAS-Alibaba Quantum Computing Laboratory, Shanghai 201315, China}
\affiliation{$^{6}$\it CAS Centre for Excellence and Synergetic Innovation Centre in Quantum Information and Quantum Physics, University of Science and Technology of China, Hefei, Anhui 230026, China}
\begin{abstract}
We experimentally investigate the quantum criticality and Tomonaga-Luttinger liquid (TLL) behavior within one-dimensional (1D) ultracold atomic gases. Based on the measured density profiles at different temperatures, the universal scaling laws of thermodynamic quantities are observed. The quantum critical regime and the relevant crossover temperatures are determined through the double-peak structure of the specific heat. In the TLL regime, we obtain the Luttinger parameter by probing sound propagation. Furthermore, a characteristic power-law behavior emerges in the measured momentum distributions of the 1D ultracold gas, confirming the existence of the TLL.
\end{abstract}

\maketitle
Quantum many-body systems can exhibit phase transitions even at zero temperature \cite{Sachdev:2007,Fisher:1989}. Here, quantum fluctuations arsing from Heisenberg's uncertainty relation drive the transition from one phase to another. In this regard, one-dimensional (1D) quantum systems are special owing to the significant microscopic fluctuations which induce a continuous phase transition between a disordered state and a TLL \cite{Haldane:1981,Giamarchi:2004,Zhou:2010,Cazalilla:2011,Guan:2013a}. Near the transition point, a quantum critical regime emerges at finite temperatures and separate these two phases \cite{Sachdev:2007,Fisher:1989,Sachdev:2011}. Although the 1D low-energy physics is generally described by the well-established TLL theory \cite{Haldane:1981}, experimental investigations of the TLL and its related quantum criticality are rare \cite{Lake:2005,Rueegg:2008,Kono:2015}. In this context, signatures of TLL were found in some 1D systems, such as organic conductors \cite{Schwartz:1998}, carbon nanotubes \cite{Yao:1999}, spin ladders \cite{Rueegg:2008}, and quantum gases \cite{Paredes:2004,Hofferberth:2008}. Among these strongly correlated systems, ultracold atomic gases offer a great precision and tunability for studying quantum phase transitions \cite{Greiner:2002,Bloch:2008} and critical phenomena \cite{Donner:2007,Zhang:2012}. However, observation of quantum criticality and determination of the TLL boundary in 1D quantum gases remain elusive.

\begin{figure}[!t]
\centerline{\includegraphics[width=8cm]{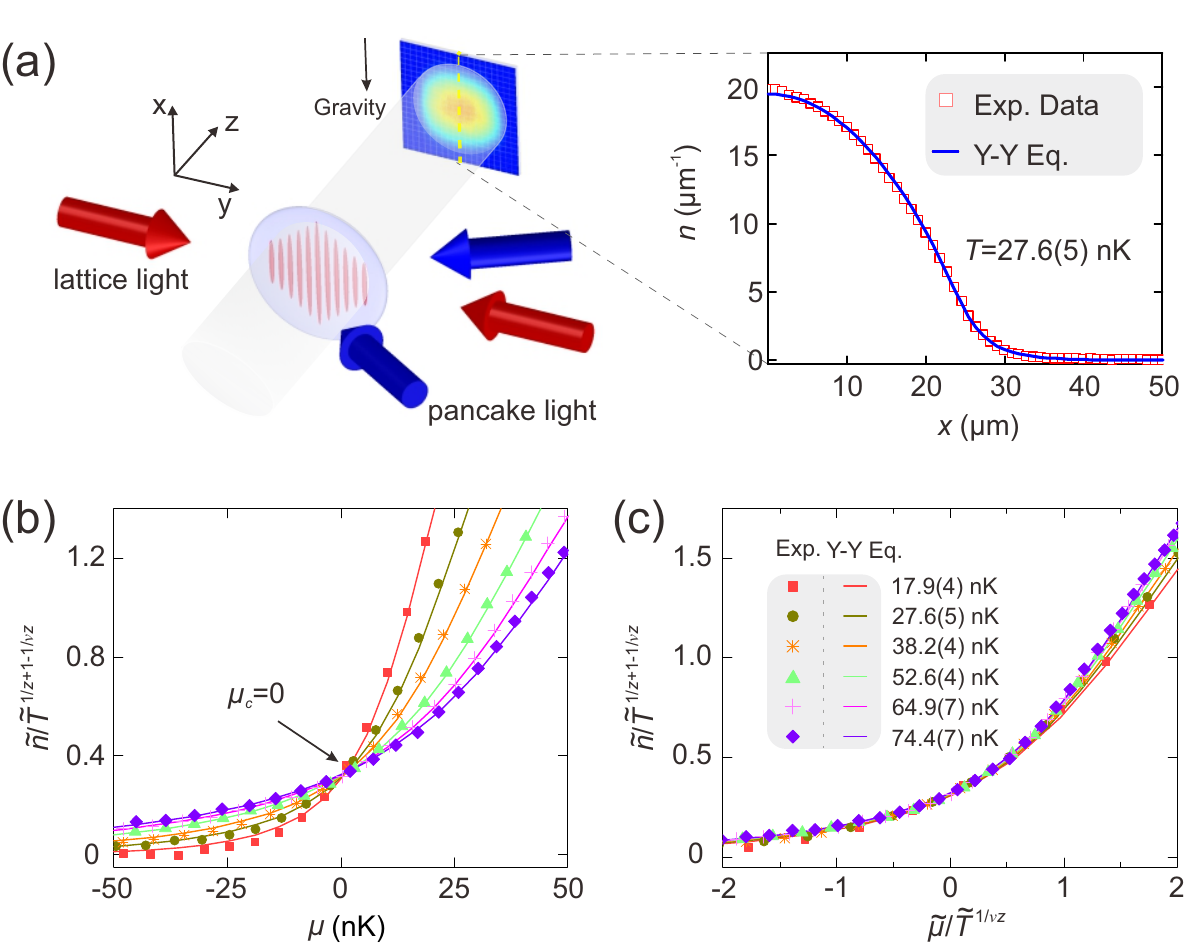}}
\caption{Experimental setup and density scaling law. (a)  The 1D system consists of an array of tubes  created by a blue-detuned ``pancake" lattice and a red-detuned retro-reflected lattice. The  density profiles are measured by {\em in situ}  absorption imaging. The inset shows an average line-density in comparison with the prediction of the Y-Y equation.  (b) The rescaled densities at different temperatures intersect at the critical point $\mu_c=0$. Here $\tilde{n}=n/c$, $\tilde{\mu} =\mu/(\hbar^2c^2/2m)$ and $\tilde{T}=k_BT/(\hbar^2c^2/2m)$, with $c=-2/a_{1D}$. The symbols denote the experimental data, solid curves stand for the theoretical predictions. (c) At different temperatures, the rescaled densities against  $\tilde{\mu}/\tilde{T}^{1/\nu z}$ collapse into a single curve around $\mu_c$.} \label{fig:fig1}
\end{figure}

In this Letter, we report the observation of quantum criticality and evidence of TLL in 1D ultracold Bose gases of $^{87}$Rb. The atomic samples at different temperatures are prepared in well-designed 1D harmonic potentials. Using a high-resolution microscope, we measure the density profiles by {\em in situ} absorption imaging. The density scaling law is obtained by rescaling these measurements at different temperatures and chemical potentials. Based on the thermodynamic relations \cite{Ho:2010,Yang:1969,Guan:2011}, we derive the pressures and entropy densities, which exhibit similar universal scaling around the critical point. Moreover, we determine two crossover branches that distinguish the quantum critical (QC) regime from the classical gas (CG) and the TLL through the double-peak structure of the specific heat. To further investigate the degenerate gas, we probe the propagations of density disturbances and acquire the Luttinger parameters. Then we characterize the phase correlation of the 1D ultracold gas through its momentum distribution. According to the bosonization-based theory \cite{Haldane:1981,Cazalilla:2004}, the obtained power-law behavior in the momentum profiles confirms the existence of the TLL.

The experiment starts by adiabatically loading a Bose-Einstein condensate of $\SI{\sim2}\times10^5$ atoms into a single layer of a pancake-shaped trap. We then confine the atoms into an array of isolated tube-shaped traps arranged in a plane by superimposing another red-detuned lattice with wavelength $\lambda_r = 1534$ nm into the system [see Fig.~\ref{fig:fig1}(a)]. Owing to the homogeneity of the light beams among these tubes, they are identical to each other with trap frequencies $\omega_x = 2\pi \times 22.2(1)$ Hz and $\omega_{\perp} = \sqrt{\omega_y \omega_z} = 2\pi \times 7.99(1)$ kHz. The spatial resolution of the imaging system ($1.0\ \mu \text{m}$) is slightly larger than the lattice spacing $\lambda_r/2 = 767$ nm. After acquiring around 400 high-resolution images for each experimental setting, we then obtained very precise 1D density profiles by averaging these images. The measured temperatures \cite{supp}, $T = 18\ $-$\ 74$ nK, and corresponding chemical potentials in the trap center, $\mu_0 = 67\ $-$\ 93$ nK, satisfy the 1D conditions of $k_\text{B}T,\, \mu_0\ll\hbar\omega_\perp$. The minimal entropy per particle of $0.055(1)$ $k_\text{B}$ at $T=17.9(4)$ nK and $\mu_0=67.1(1)$ nK indicates that the 1D gas is strongly degenerate. The dimensionless  interaction parameter $\gamma \approx 2 m a_{1D} \omega_{\perp}/(\hbar n_0)= 0.04\ll1$ suggests that the central region of the system is in the weakly interacting regime, where $m$ is the mass of atom, $a_{1D}$  is the 1D effective scattering length and $n_0 $ is the line density at the center of the 1D tube. Under such experimental conditions, the 1D Bose gas can be described by the Lieb-Liniger model \cite{Lieb:1963}. Within the local density approximation (LDA), the measured densities agree well with the theoretical predictions from the Yang-Yang (Y-Y) exact grand canonical theory \cite{Yang:1969} [see Fig.\ref{fig:fig1}(a)].

\begin{figure}[!t]
\centerline{\includegraphics[width=8.4cm]{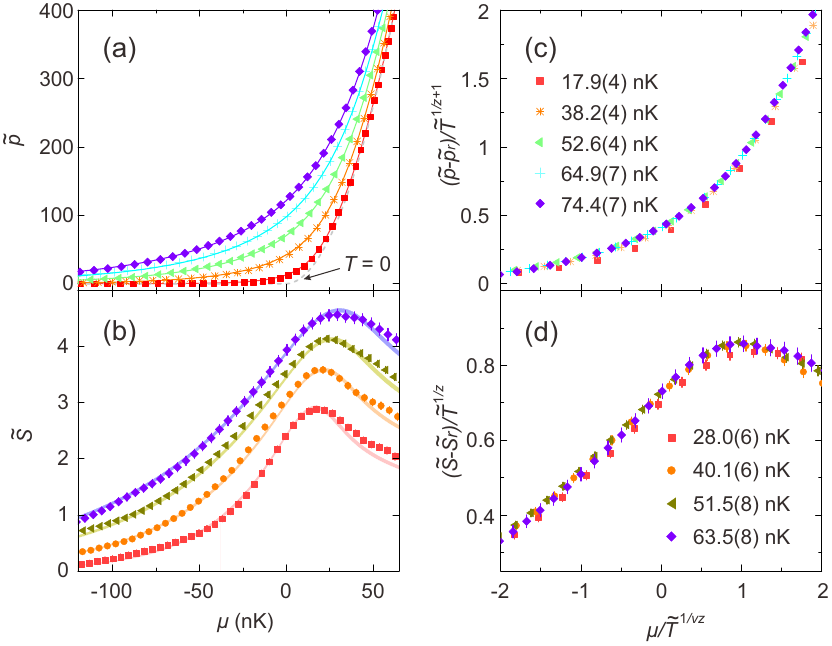}}
\caption{The dimensionless pressure $\tilde{p}=p/[\hbar^2 c^3 /(2m)]$ and entropy density $\tilde{S}=S/(k_Bc)$. The solid lines and the shaded curves are the theoretical predictions from the Y-Y equation. The thickness of the shaded area indicates errors arising from the temperature uncertainties. (a) The pressure EOS  (symbols) at different temperatures are deduced from the density profiles. (b) Symbols denote the experimental data of entropy densities extracted from the pressure. (c) and (d), Using the same critical exponents $\nu$ and $z$ for the density, the rescaled pressures and entropy densities overlap and collapse into a single curve around the critical point. Here $p_r$ and $S_r$ are the regular parts of the scaling functions. The error bars denote the $\pm 1\sigma$ statistical errors.} \label{fig:fig2}
\end{figure}

For the one-dimensional Lieb-Liniger model \cite{Guan:2011,Lieb:1963} at zero temperature, a vacuum-to-TLL phase transition occurs when we change the chemical potential in a positive direction across the critical point $\mu_c=0$.  At finite temperatures, a QC regime emerges near $\mu_c$ and separates the CG and the TLL phase. In the QC regime, the correlation length $\xi$ diverges as $\xi \propto |\mu-\mu_c|^{-\nu}$, and the energy gap $\Delta$ is inversely proportional to the correlation length $\Delta \propto \xi^{-z} \propto |\mu-\mu_c|^{\nu z}$, which vanishes as $\mu \rightarrow \mu_c$ \cite{Sachdev:2007,Fisher:1989,Guan:2011}. Here $\nu$ and $z$ are defined as the correlation length exponent and the dynamic critical exponent, respectively. In this context, the particle density in QC obeys a universal scaling law, as $n(\mu,T) =T^{\frac{d}{z}+1-\frac{1}{\nu z}}{\cal{F}}\left(\frac{\mu-\mu_c}{T^{\frac{1}{\nu z}}}\right)$, where the dimensionality is $d=1$ and ${\cal{F}}(x)$ is the scaling function \cite{Zhou:2010}.

Such universal scaling law is extracted from the density profiles at temperatures ranging from 17.9(4) nK to 74.4(7) nK. As shown in Fig.~\ref{fig:fig1}(b), we identify the critical point using that the scaled density becomes temperature-independent at $\mu_c$, i.e., the density profiles at different temperatures intersect at the critical point. The critical exponents $\nu$ and $z$ are determined by the overlapping feature of the rescaled density profiles \cite{supp}. The rescaled measurements fall into a single curve with $\nu=0.56^{+0.07}_{-0.08}$ and $z= 2.3^{+0.6}_{-0.3}$ [Fig.~\ref{fig:fig1}(c)], confirming the emergence of the quantum critical scaling. Here the uncertainties correspond to a 95$\%$ confidence level. The critical exponents agree with the predictions from the Y-Y equation, $\nu=0.5$ and $z =2$  \cite{Yang:1969,Guan:2011}. The above properties of densities at various temperatures and chemical potentials reveal the nature of scaling invariance.

The thermodynamics of the 1D system at equilibrium are described by the equation of state (EOS). We can derive the local pressure EOS from the atomic density via $p(\mu, T) = \int_{-\infty}^{\mu}n(\mu', T)d\mu'$ \cite{Ho:2010} by introducing a proper cut-off in the CG regime, as shown in Fig.~\ref{fig:fig2}(a). For the lowest temperature experimentally probed $T=17.9(4)$ nK, the population below $\mu_c$ is negligible and therefore the pressure approaches that of the zero temperature result. Whereas at higher temperatures, the pressure curves split clearly in the QC regime and bunch up again at large chemical potentials. From the pressure EOS, one can obtain  other thermodynamic properties. For example, the entropy density can be deduced as, $S(\mu,T) = \left[\partial p(\mu,T) /\partial T\right]_{\mu}$ \cite{Ho:2010}. Fig.\ref{fig:fig2}(b) shows the entropy densities extracted from experimental data and the theoretical curves. Peaks arise in the entropy density curves and become flatter at higher temperatures, revealing enhanced disorder in the QC regime. Moreover, both the pressure  and entropy densities [Fig.~\ref{fig:fig2}(c)(d)]  have similar universal scaling laws with the same critical exponents as those in the density scaling function \cite{supp}.

\begin{figure}[!t]
\centerline{\includegraphics[width=6cm]{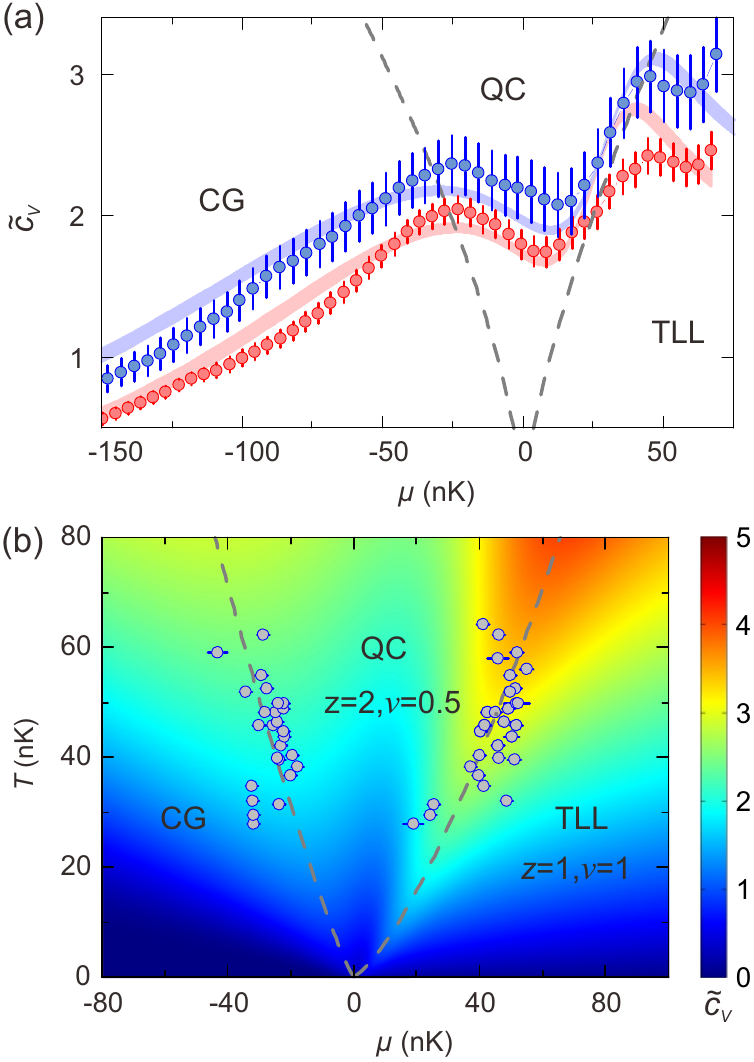}}
\caption{The dimensionless specific heat $\tilde{c}_V=c_V/(k_Bc)$ and its phase diagram. (a) Experimental specific heat (red and blue circles) at the temperatures of $T = 40(1)$ nK and $T = 50(1)$ nK, respectively. The double-peak feature of specific heat marks out the regimes CG, QC and TLL. The shaded areas indicate the theoretical predictions by taking account of the temperature uncertainty. (b) Contour plot of specific heat in $T-\mu$ plane in which its peaks (dashed lines) separate three fluctuation regimes. The QC and the TLL are classified by different critical exponents $z$ and  $\nu$. The dots denote the experimental data of the specific heat peaks which mark the two crossover temperatures $T^{*}$.   Here the error bars denote the $\pm$1$\sigma$ fitting errors.} \label{fig:fig3}
\end{figure}

In the scenario of quantum criticality, determining the crossover temperatures $T^{*}$ in quantum gases poses theoretical \cite{Cazalilla:2011,Guan:2013a} and experimental challenges \cite{Zhang:2012}. Here we make a distinction of different regimes through the feature of specific heat, $c_V=T\left( \partial ^2 p/\partial T^2 \right)_{\mu}$, which manifests different scales of the energy fluctuations in the grand canonical ensemble. In ultracold atomic gases, obtaining a $c_V$ of merit requires high precision density measurements. At finite temperatures, a double-peak structure of the specific heat appears and marks two crossover temperatures fanning out from the critical point [Fig.~\ref{fig:fig3}(a)]. The peak values of the $c_V$ constitute two branches of the QC crossover boundaries in Fig.~\ref{fig:fig3}(b). Here, the theoretical curves and contour plot of specific heat are numerically calculated via the second-order derivation of the pressure. The left branch indicates the 1D degenerate condition, i.e. the thermal de Broglie wavelength $\lambda_T=\sqrt{2\pi \hbar^2/(mk_\text{B}T)}$ is approximate to the atomic spacing $1/n$. The right branch separates the QC and a linear-dispersion TLL regime with a crossover temperature $T^{*}\sim |\mu-\mu_c|^{\nu z} $. In the TLL regime of the phase diagram, the specific heat at a certain chemical potential depends almost linearly on temperature \cite{supp}, reflecting a collective behavior of the quantum liquid. The peaks of $S$ and the valleys of $c_V$ reveal that quantum fluctuations dominate the quantum critical behavior.

\begin{figure*}[!t]
\centerline{\includegraphics[width=16.5cm]{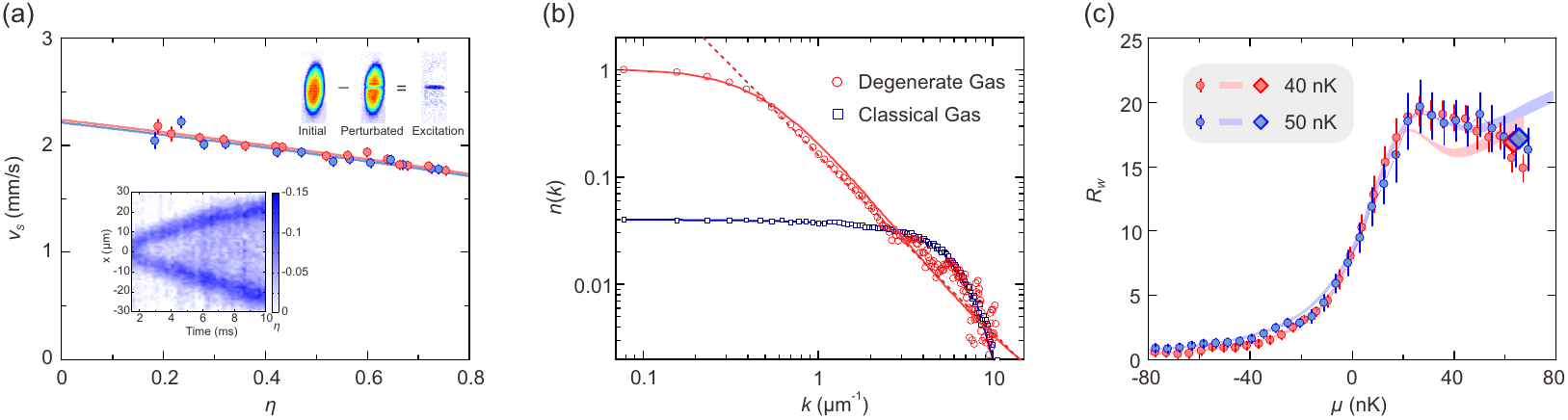}}
\caption{Evidences for TLL. (a) Measurements of sound velocity. The upper inset shows the generation process of the excitation in the sample. The excitation signal is obtained by subtracting the perturbated cloud from the initial density. Lower inset shows the propagation of a negative perturbation in 1D tubes with amplitude $\eta = 0.17(1)$ and a Gaussian width $w = 4.0(4) \mu$m. The red and blue circles are experimental sound velocities versus excitation ratios at temperature $T = 40(3)$ nK and $T = 50(3)$ nK. The red and blue curves are the fitting results. (b) Momentum profiles of the 1D gases. The red circles and blue squares stand for the experimental momentum distributions for a degenerate gas at $T=40(1)$ nK and a classical gas at $T=209(1)$ nK, respectively. All the experimental data are normalized to the zero-momentum value of the degenerate gas. The red solid curve is the theoretical prediction by considering the finite-temperature effect and the averaged Luttinger parameter. The red dash curve is an auxiliary straight line with a slope of -1.66, while the blue solid curve follows a Gaussian distribution. (c) Wilson ratio $R_W$ and Luttinger parameter $K$. The red and blue circles are experimental $R_W$ at $T=40(1)$ nK and $T=50(1)$ nK, while the shaded areas indicate the theoretical predictions. Two diamond points represent the averaged Luttinger parameter of the TLL regime. The error bars represent the $\pm 1 \sigma$ statistical errors. } \label{fig:fig4}
\end{figure*}

The low-energy properties of the TLL can be fully described by the sound velocity $v_s$ and Luttinger parameter $K$ \cite{Giamarchi:2004,Cazalilla:2011,Haldane:1981a}. Here $v_s$ represents the propagating velocity of density disturbances, which satisfies a linear dispersion relation $\omega = v_s |k|$. Experimentally, the sound velocity is obtained by monitoring the propagation of density perturbations in the 1D tubes. We apply a magnetic gradient along the longitudinal direction ($x$ axis) to create a spatially-dependent Zeeman splitting, which enables a spatially-resolved transfer of atoms from $\Ket{F=1,m_F=-1}$ into $\Ket{F=2,m_F=0}$ sublevel via microwave (MW) transitions. With a resonant light pulse to remove the atoms in the $\Ket{F=2}$ states, density dips are generated in the center of the 1D tubes. The profile of these defects is approximately Gaussian $\eta n_0 e^{-x^2/2 w^2}$, where the relative amplitude $\eta$ and the width $w$ are tailored by adjusting the MW strength. As shown in the insets of Fig.~\ref{fig:fig4}(a), such negative perturbations split into two parts and then symmetrically propagate along the 1D tubes. For different perturbing amplitudes, we resolve a linear relation between the $v_s$ and the square root of remaining density $\sqrt{n_0 (1-\eta/2)}$, as $v_s (\eta)= v_s(0)\sqrt{1-\eta/2}$ \cite{Andrews:1997,Kavoulakis:1998,Meppelink:2009}. Based on this relation, the sound velocities at vanishing perturbations are determined as $2.24(1)$ mm/s and $2.21(1) $ mm/s for $T=40(3)$ nK and $T=50(3)$ nK, respectively. For a uniform quantum gas, the Luttinger parameter $K$ and the $v_s$ have a relation as $K = \hbar \pi n/m v_s$ \cite{supp,Haller:2010}. Whereas for TLL in the harmonic trap, we can get an averaged Luttinger parameter $\bar{K}$ by employing the averaged density and sound velocity over the TLL regime. In the cases of $T=40(3)$ nK and $T=50(3)$ nK, the $\bar{K}$ acquired from the measured sound velocities and atomic densities are $16.9$ and $17.2$, respectively \cite{supp}.

For the TLL, one characteristic property of the TLL is the interaction-dependent power-law behavior of correlation functions \cite{Haldane:1981,Haldane:1981a,Cazalilla:2004,Giamarchi:2004}. Such quasi-long-range order is evident in the first-order correlation function $g^{(1)}(x)=\langle \psi^{\dagger}(x)\psi(0) \rangle$, which features a power-law decay $g^{(1)}(x) \propto x^{-1/2K}$ in the uniform system at zero temperature \cite{Haldane:1981}. The momentum distribution of TLL is the Fourier transform of this correlation function, i.e. $n(k) = \int_{-\infty}^{\infty} \text{d}x e^{-\text{i}kx}g^{(1)}(x)$. In a low-temperature atomic gas, the dominant phase fluctuations give rise to a finite correlation length $l_{\phi} = \hbar v_s K/(\pi k_B T)$ \cite{Cazalilla:2004,Bloch:2008} and modify the power-law behavior of the momentum distribution. In this case, $n(k)\simeq A(K) \text{Re}[\Gamma(1/4K + \text{i}kl_{\phi}/2K)/\Gamma(1-1/4K+\text{i}kl_{\phi}/2K)]$, where $A(K)$ is a $K$-dependent parameter \cite{Cazalilla:2004}. If the system size $L$ is much larger than the phase correlation length $l_{\phi}$, the inhomogeneity of the harmonic trap can be safely neglected and the system can be treated using the LDA \cite{Cazalilla:2004,Richard:2003,Jacqmin:2012}.

To access the momentum distribution of the 1D gases, we utilize a focusing technique during the time-of-flight \cite{Shvarchuck:2002} instead of a conventional long-time expansion \cite{supp}. The momentum distribution of a 1D gase at $T=40(1)$ nK is displayed by a log-log plot in Fig.\ref{fig:fig4}(b). For this non-uniform system, the averaged Luttinger parameter of the TLL regime ($\mu \geq T^*$) is $\bar{K} =15.9$, indicating a correlation length of $l_{\phi} = 1.9 \ \mu$m. Thermal fluctuations break the long-range phase correlations, making the momentum distribution for $k < 1/l_{\phi}$ rather flat. Another characteristic length is the healing length $\sim 0.1\ \mu$m, which determines the high-momentum cut-off of our measurements. The measured $n(k)$ exhibits a power-law decay at intermediate momenta with a linear slope of $-1.66$ ($1/l_{\phi} \leq k \leq 20/l_{\phi}$). As the system satisfies the condition of LDA $\left(L \gg l_{\phi}\right)$, we obtain a theoretical curve of $n(k)$ by using the parameter $\bar{K}$. This curve has an asymptotic power-law decay with the slope $-1+1/2K$ at large momenta ($k > 40/l_{\phi}$) \cite{Cazalilla:2004,supp}. The inhomogeneity of the harmonic potential might lead to some modification to $n(k)$, which would have a extended flat region at small momenta and a Lorentzian distribution with power-law exponent -2 at intermediate momenta \cite{Richard:2003,Jacqmin:2012}. However, within the accessible range, our experimental result agrees well with the theoretical prediction \cite{Cazalilla:2004}, indicating that TLL behavior dominates the system and the momentum distribution can be qualitatively understood by considering a uniform gas with the same $\bar{K}$. For a comparison, we also measure the momentum distribution of a classical gas with $T=209(1)$ nK and $\mu_0 = -111(1)$ nK. Both the spatial and the momentum distribution of this gas show classical Gaussian profiles as predicted by the Boltzmann distributions.

In the TLL regime, although the Fermi liquid theory cannot describe 1D systems due to collective behavior herein, two important features of quantum liquid still retain, i.e. the compressibility $\kappa$ is independent of temperature and the specific heat $c_V$ is linearly proportional to temperature \cite{Ninios:2012,Guan:2013,Schofield:1999}. Therefore, we employ a dimensionless Wilson ratio to characterize different regimes, $R_W = \pi^2 k_B^2/3 \cdot \kappa/(c_V/T)$ \cite{Wilson:1975,Ninios:2012,Guan:2013,Yu:2016}. An equivalence between the Wilson ratio and the Luttinger parameter has been proved in the uniform TLLs \cite{Guan:2013,Yu:2016}. This relation indicates that the particle number fluctuation and the energy fluctuation are on an equal footing with respect to $T$. In our experiment, the derived $R_W$ approaches the averaged $K$ in the TLL regime [see Fig.\ref{fig:fig4}(c)]. The connection between the $R_W$ and $K$ provides a novel method for determining the Luttinger parameter in the solid-state system \cite{Ninios:2012}, where the sound velocity is hard to measure. Meanwhile, the crossover features of the QC regime can also be characterized by the ``critical cone'' in the phase diagram of $R_W$ \cite{supp}.

In summary, we present a systematic study of the quantum criticality and TLL behavior in 1D quantum gases. The 1D density profiles of ultracold Bose gases with a minimum entropy per particle $0.055\, k_B$ have been obtained with a high precision. Using these density profiles, we have determined universal scaling laws, the EOS and crossover temperatures of this system. Afterwards, the Luttinger parameters have been obtained by the measured sound velocities and atomic densities. In our non-uniform system, the momentum distribution which exhibits a power-law decay at intermediate momenta is well consistent with the TLL theory. Our experiment provides prototypical methods for studying quantum critical phenomena and quantum liquids, not only in other spinless quantum gases \cite{Kinoshita:2004,Paredes:2004,Jacqmin:2011,Hofferberth:2008,Haller:2010} but also in quantum many-body systems involving rich spin (and charge) interactions and symmetries, such as spin chains \cite{Kono:2015,Pagano:2014}, the Yang-Gaudin model \cite{Liao:2010} and the Hubbard model \cite{Essler:2005}.

The authors would like to thank H Zhai, Q. Zhou,  A. del Campo, S. Jochim, S. Whitlock, Y.-J. Deng, Z.-C. Yan, A. Truscott and T. Giamarchi for helpful discussions. This work has been supported by the National Key R\&D Program of China (Grant Numbers  2016YFA0301600, 2017YFA0304500), the NNSFC (Numbers. 11534014,  11374331, 91221204) and the Chinese Academy of Sciences. XWG  has been partially supported by the Australian Research Council.

{$^*$These authors contributed equally to this work.}\\
{\indent $^\dagger$Email: xiwen.guan@anu.edu.au, yuanzs@ustc.edu.cn, or pan@ustc.edu.cn}

\bibliographystyle{apsrev4-1}
\bibliography{QCLD}
\clearpage
\setcounter{page}{1}
\def\scr{\mathscr}
\def\SG{{\scr G}}
\setcounter{equation}{0}
\setcounter{figure}{0}
\makeatletter
\makeatother
\global\def\theequation{S\arabic{equation}}
\global\def\thefigure{S\arabic{figure}}
\onecolumngrid
\textbf{\large\noindent Supplemental Material for ``Observation of quantum criticality and Tomonaga-Luttinger liquid in one-dimensional Bose gases'' B. Yang et al\\ \vspace{1cm}}
\twocolumngrid

\section{One-dimensional system and the calibrations}
\begin{figure}[!b]
\centering{\includegraphics[width=8.5cm]{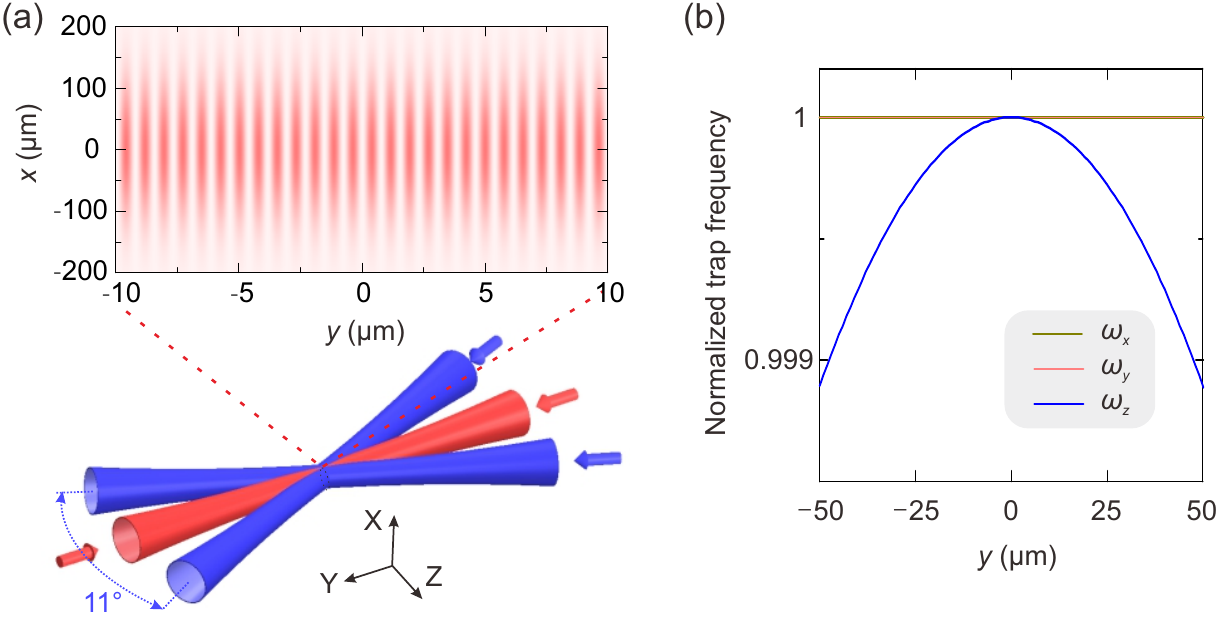}} \caption{Schematic of the 1D tubes and estimation of the homogeneity. (a) Illustration of the 1D system, the pancake trap and the red lattice form an array of 1D tubes. (b) The inhomogeneity among the tubes is characterized by the trap frequencies $\omega_{x,y,z}(0,y,0)$. Among the 1D tubes within the atomic cloud size ($\pm 30 \mu$m along $y$), the trap frequencies have negligible change.} \label{figS0}
\end{figure}

The experiment starts by loading a $^{87}$Rb Bose-Einstein condensate of $\sim2\times10^5$ atom into a 2D optical lattice. The condensate is first compressed along the $z$ direction and then adiabatically loaded into a single layer of a pancake lattice. This pancake lattice is generated by interfering two coherent $\lambda_b = 767$ nm laser beam with $11^{\circ}$ of intersection angle. In such pancake potential, we hold the atoms for an equilibration time of 3 seconds and then the atom number decreases to $\sim 4\times 10^4$ due to three-body loss and further evaporation. At this stage, the temperature of the Bose gas can be well controlled. Afterwards, we superimpose another red-detuned lattice at wavelength $\lambda_r = 1534$ nm into the system, forming an array of 1D tubes. Both the pancake lattice and red lattice are then ramped up in 200 ms to reach a highly anisotropy trapping potential. In the final 1D condition, the red lattice reaches 27.8(1)$E_r$ to well isolate atom tunnelling (tunnelling rate $0.6$ Hz) among lattice sites, here the $E_r \!=\! k_B \times 47$ nK denote the recoil energy. The geometry of the laser beams is illustrated in Fig. \ref{figS0}(a), where the 1D tubes are aligned in the $y$ direction. The beam waist of the pancake and red-lattice light are 144 $\mu$m and 155 $\mu$m, respectively. The Rayleigh lengths of these beams are much larger than the cloud size along $y$. As shown in Fig. \ref{figS0}(b), the inhomogeneity of the trap frequencies is negligibly small, indicating all the tubes are identical.

\begin{figure}[!t]
\centering{\includegraphics[width=8.5cm]{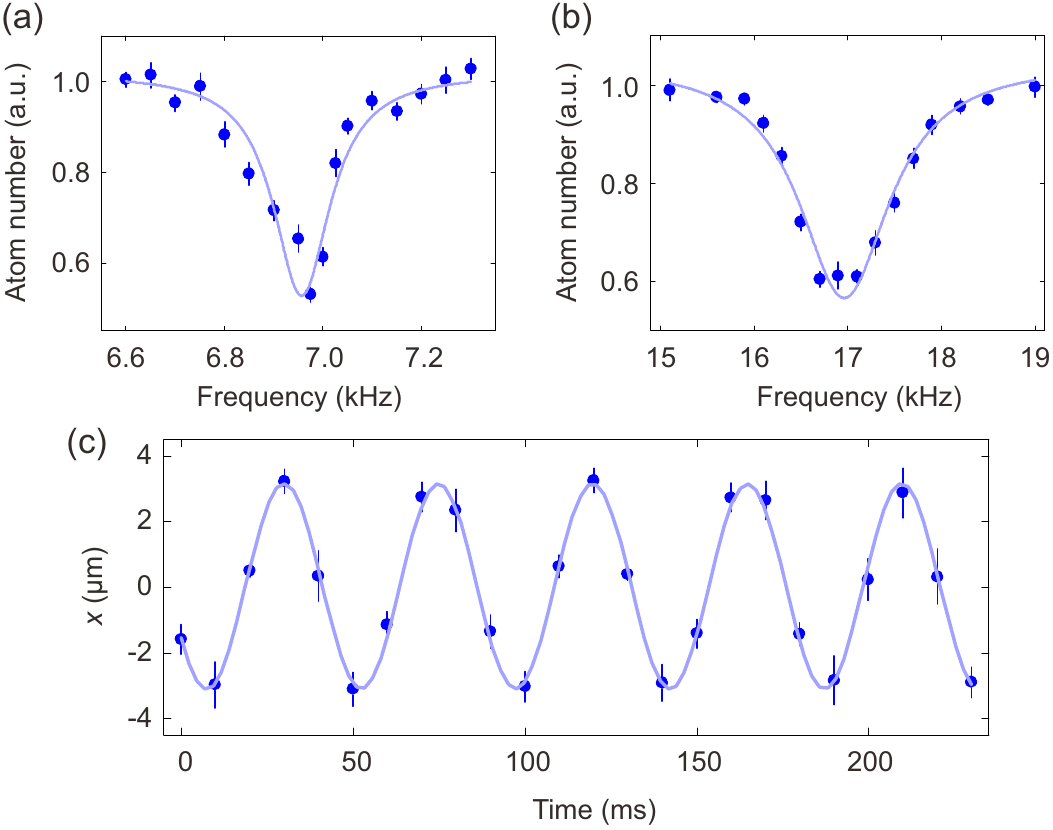}} \caption{Calibration of the 1D system. (a), (b) The excitation spectrum of the pancake trap and the red lattice. The atom numbers are normalized to the unperturbed 1D conditions.  The spectra are fitted by Lorentzian functions. (c) Dipole oscillation along $x$. In these figures, the error bars indicate the standard deviations of the measurements.} \label{figS1}
\end{figure}

The trap frequencies of our 1D system are calibrated via the excitation spectroscopy of the ultracold gases. First, a cloud sample is prepared at thermal equilibrium in the synthetic lattice potential. To calibrate the frequency along $z$, we then apply a sinusoidal amplitude modulation to the pancake lattice at varying frequency with about 10$\%$ of the trap depth for 200 ms. When the modulation frequency matches the trap frequency, atoms can be excited to higher vibrational levels and then escape from the trap. By probing the residual atom number, we obtain a spectrum as Fig. \ref{figS1}(a). The resonance frequency of $z$ direction is $\omega_z = 2\pi \times 6.96(2)$ kHz. A similar modulation scheme is implemented to calibrate the trapping along $y$. The residual atom number is sensitive to the modulation when its frequency approaches the energy gap between the ground and second excited band. As shown in Fig. \ref{figS1}(b), the 16.96(3) kHz band gap indicates that the trap frequency along $y$ is $\omega_y = 2\pi \times 9.17(2)$ kHz. The axial confinement is calibrated by measuring the frequency of dipole oscillation. We apply a potential displacement to the 1D tubes along $x$ and then monitor the center-of-mass of the bulk. The dipole oscillation frequency equals to the trap frequency, which is resolved from Fig. \ref{figS1}(c) as $\omega_x = 2\pi \times 22.2(1)$ Hz.

To measure the atomic density precisely, the general Beer-Lambert law needs to be modified when we consider the highly saturated imaging, the stray magnetic field and properties (polarization or line-width) of the probing laser. To model these effects, we use two scaling parameters $\alpha$ and $\beta$ for calibrating the absorption imaging. The modified Beer-Lambert law \cite{Yefsah:2011} is,
\begin{equation}
n(x,y)\sigma_0 = \frac{1}{\beta}\left[ -\alpha \ln\frac{I_f(x,y)}{I_i(x,y)} + \frac{I_i(x,y)-I_f(x,y)}{I_{sat}} \right]. \label{Imaging}
\end{equation}
Here, $n(x,y)$ is the density distribution, $\sigma_0$ is the scattering cross section of D2-line cycling transition for circular polarized light, $I_{sat}$ is the saturation intensity, $I_i(x,y)$ and $I_f(x,y)$ are intensity distributions of the probe beam before and after absorption. The parameter $\alpha$ is calibrated by applying different intensities of the probe beam to a thermal cloud while preserving the same value of measured density \cite{Reinaudi:2007,Dai:2016}. The parameter $\beta$ represents a linear shift of the absolute optical depth. Its calibration relies on the Thomas-Fermi distribution of an ultracold two-dimensional (2D) quantum gas. The trapping frequencies of the 2D gas can be measured very precisely. Then we find there only exists one self-consistent solution for $\beta$ in determining the atomic density. These calibrations give $\alpha =2.6(2)$ and $\beta = 0.872(3)$. The density profiles of 1D gases are obtained by illuminating the cloud with a strong probing beam ($\SI{\sim100}I_{sat}$), which reduces the error arising from the $\alpha$ term and leads to a small calibration error (0.4$\%$) of the atomic density.

\section{Thermometry of the 1D gases}
During the adiabatic loading process, the lattice potentials are turned on slowly to avoid excitations of dipole oscillations and breathing modes. After the ramping, the center-of-mass and the size of the atomic cloud are measured. The amplitudes of dipole oscillation ($\sim 1\ \mu$m) and breathing mode ($\sim 1\%$ of the half-width) are fairly small. Such an adiabatic process produces arrays of isolated 1D systems in thermal equilibrium. After a holding time of 250 ms in the 1D traps, we measure the \emph{in situ} density distribution of the cloud by performing an absorption imaging with a probe light propagating along $z$. Our imaging objective has a numerical aperture of 0.48, providing a high spatial resolution for measuring the density distribution. Each pixel on the CCD camera corresponds to an area of $0.9\times 0.9\ \mu \text{m}^2$ on the atom plane. The effective pixel size is comparable to the separation of neighbouring tubes $\lambda_r/2 = 767$ nm.

To precisely measure the mean density of the 1D gases, we acquire a large number of images (typically 400) taken in the same experimental condition. The technical noise and fluctuations can be removed by averaging the measurements. Averaging over plenty of samples $N_s$ can reduce the density variance, which converges as $\text{Var}(n)/N_{s}$. Since each 1D tube has the same trapping condition, we sort the 1D systems into groups by particle number per tube (with $2.7\%$ deviation of atom number $N$ within tubes) and then average the densities among each group.

At thermal equilibrium, the thermodynamics of the 1D gas can be fully described by the Yang-Yang equation [21].
\begin{equation}
\varepsilon(k)= \frac{\hbar^2 k^2}{2m} -\mu- \frac{k_B T c}{\pi} \int_{-\infty}^{\infty}  \frac{dq\,\ln\left(1+{\mathrm e}^{-\frac{\varepsilon(q)}{k_BT}}\right)}{c^2+(k-q)^2}, \label{TBA}
\end{equation}
where $\varepsilon(k)$ is the dressed energy, $k$ is quasi-momentum. $c=-2/a_{1D}$ is the  interaction parameter that determined by the 1D effective scattering length $a_{1D} = \left(-a_{\bot}^2/2a_s\right)\left(1-1.46 a_s/a_{\bot}\right)$ . Here, $a_{\bot} = \sqrt{2\hbar/m\omega_{\bot}}$ and $a_s$ is the three-dimensional scattering length. With a given chemical potential $\mu$ and temperature $T$, we can numerically solve this equation and acquire the dressed energy $\varepsilon(k)$. The pressure is related to the $\varepsilon(k)$ via $p=k_B T \int_{-\infty}^\infty dk \, \ln(1+\mathrm{e}^{-{\varepsilon(k)}/{k_BT}})/2\pi$, from which we can further derive the atomic density by $n = \partial p/\partial \mu$ [21, 22].

\begin{figure}[!t]
\centerline{\includegraphics[width=8.5cm]{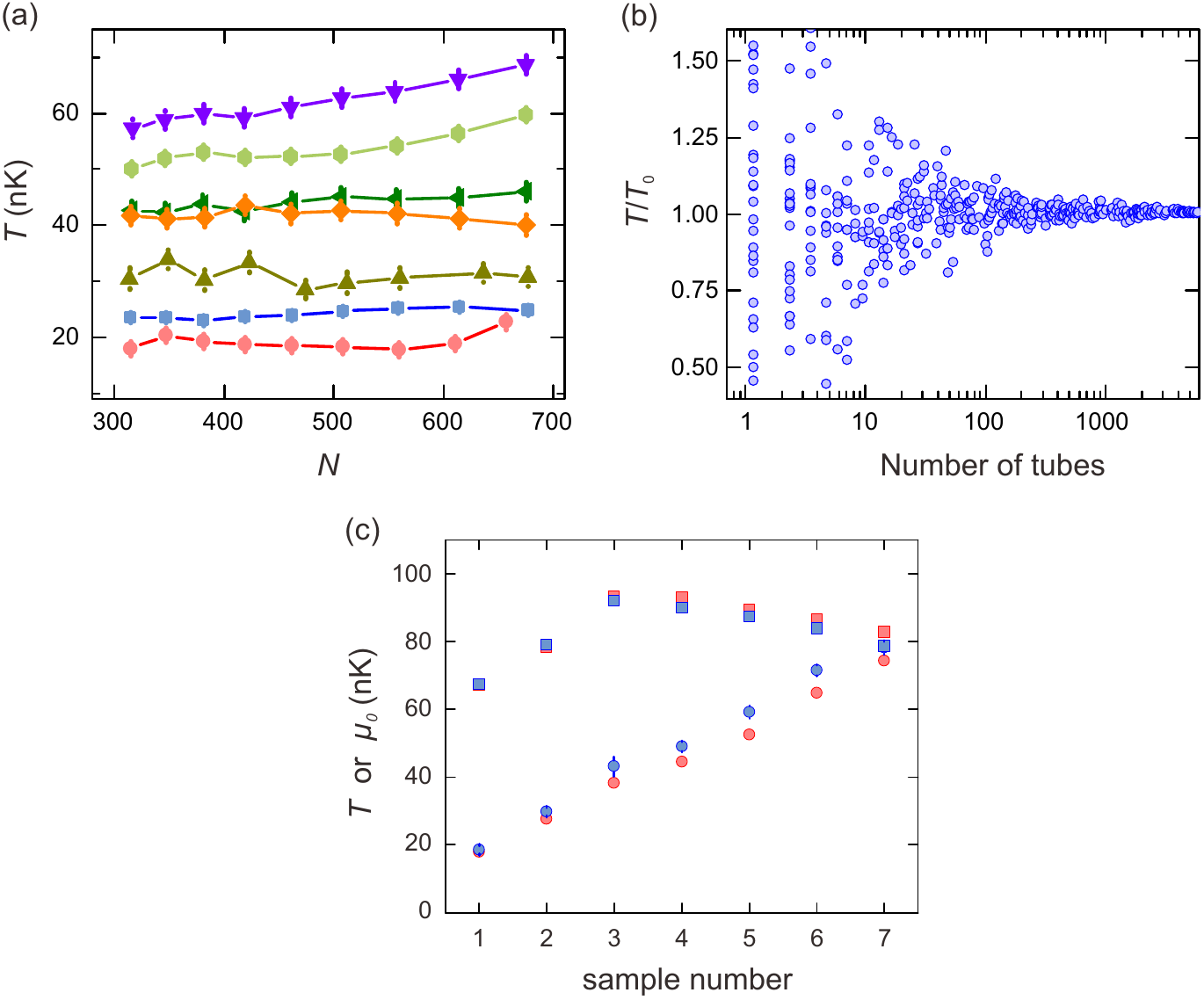}}\caption{Thermometry of the 1D gases. (a) Temperature versus atom number. Here the colours of data indicate different experimental conditions. (b) Fitting temperatures versus the number of tubes. The tubes are grouped by the atom number $N$. For one specific $N$, the fitting temperature converges to the final value $T = 38.2(4)$ nK as the increasing of the tubes samples. (c) Comparison between independent thermometry methods based on 7 groups of samples. The red points correspond to the fitting results with the Y-Y equation, in which squares and circles represent chemical potentials $\mu_0$ and temperatures $T$, respectively. The blue squares are the fitting results of chemical potentials based on Thomas-Fermi approximation. The blue circles show the temperatures extracted from the thermal wings of each density profile. } \label{FigS2}
\end{figure}

Within the local density approximation (LDA) [19, 20, 23, 31, 32], the chemical potential in  Eq. (\ref{TBA}) is replaced by the local chemical potential $ \mu \left( x \right) = \mu_0 - V \left( x \right)$. Here $\mu_0$ represent the chemical potential of trap center and $V \left( x \right) = m \omega_x^2 x^2/ 2$ is in a harmonic form. With the Y-Y equation, density distribution $n(x)$ can be obtained with a given $\mu_0$ and $T$, and vice versa. By utilizing an iteration method, we are able to find the fitting values and errors of $\mu_0$, $T$ with trap frequencies $\omega_x$, $\omega_{\perp}$ and density profile $n(x)$. Fig. \ref{FigS2}(a) shows the thermometry of the 1D system under different experimental settings. The temperatures at varying atom number $N$, i.e. different tubes, almost remain steady in each experimental condition, indicating the thermal equilibrium has been reached among and within the tubes. This should be achieved by the exchange of particles via tunnelling during the lattice ramping stage. From shot-to-shot measurements, our atom clouds have good repetition and almost no discernible drifts. As shown in Fig. \ref{FigS2}(b), we fit the temperature of the grouped tubes and find 10 images ($\sim 100$ tubes) can suppress the temperature uncertainty to be less than $10\%$.

The chemical potentials $\mu_0$ and temperatures of the 1D gases are further crosschecked with methods other than fitting the data with the Y-Y equation. The central part of the 1D gases are approximately described by the Thomas-Fermi distribution, from which the chemical potentials can be deduced through the density profiles by taking account of the weak interactions. In contrast to the highly degenerated Bose gases at the trap center, the outer wings of the cloud are in the classic gas regime, which can be considered as ideal gases. We fit the temperatures via a fugacity analysis by choosing the thermal wings of each sample. As shown in Fig. \ref{FigS2}(c), The results from these two independent methods are consistent with the thermometry with the Y-Y equation.

\section{Thermodynamics}
\begin{figure}[!htb]
\centerline{\includegraphics[width=7cm]{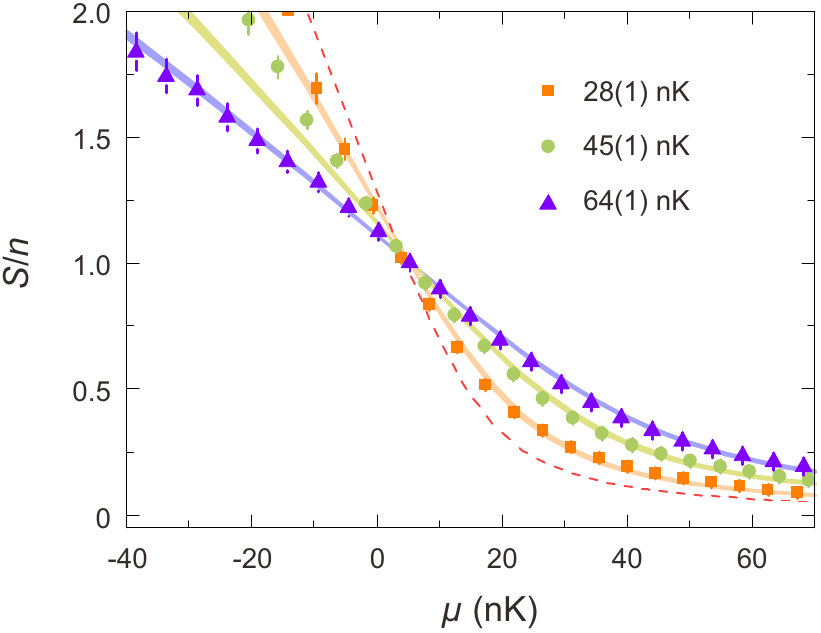}}\caption{Entropy per particle. The symbols are the data of $S/n$ at different temperatures. The error bars denote the $\pm 1\sigma$ statistical errors. The shaded regions are theoretical predictions by considering the temperature uncertainty. The dashed red line is a theoretical curve of $S/n$ at 17.9 nK, which serves as the lowest temperature of our measurements.} \label{Fig.S3}
\end{figure}

From the measured density profiles, we are able to investigate the thermodynamics of the 1D Bose gases. For the local thermal equilibrated system, the pressure $p$ and entropy $S$ can be derived based on the Gibbs-Duhem relation $dp = nd\mu + SdT$. The pressure EOS is an integral of density with respect to the chemical potential [5, 20].
\begin{equation}
p(\mu_i, T) = p_0(\mu_N, T)+\int_{\mu_N}^{\mu_i}n(\mu', T)d\mu', \label{pressure}
\end{equation}
Here $\mu_N$ denotes the cut-off of this integration, and below $p_0$ (normally, $\mu_N/T \leq -3$) the pressure can be modeled by the ideal gas. The integration of density can be well approximated by discrete summation based on the trapezoidal rules. Meanwhile, the errors of densities propagate to the pressure. The variance of pressure is in the form as $\text{Var}\left[p(\mu_i,T)\right] = \sum\limits_{j=N}^{i} \text{Var}(n_{j}) \Delta \mu_j^2$. The entropy density $S(\mu,T)$ is the differential of pressure versus temperature [20].
\begin{equation}
S(\mu,T)= \left(\frac{\partial p}{\partial T}\right)_{\mu,c},
\end{equation}
The partial derivative of pressure in the entropy function can be approximated by the discrete differential of pressure. In general, we choose two pressure curves with temperature difference of $\Delta T =20 $ nK, the entropy is calculated as $S(\mu,T)= [p(\mu,T+\Delta T/2)-p(\mu,T-\Delta T/2)]/\Delta T$. From the entropy density $S$ and the line density $n$, we get the entropy per particle in unit of $k_B$ through $S/n$. Fig. \ref{Fig.S3} shows the $S/n$ at different temperatures, from which we know the lowest entropy per particle in our measurements is 0.055(1)$k_B$.

\begin{figure}[!htb]
{\includegraphics[width=6cm]{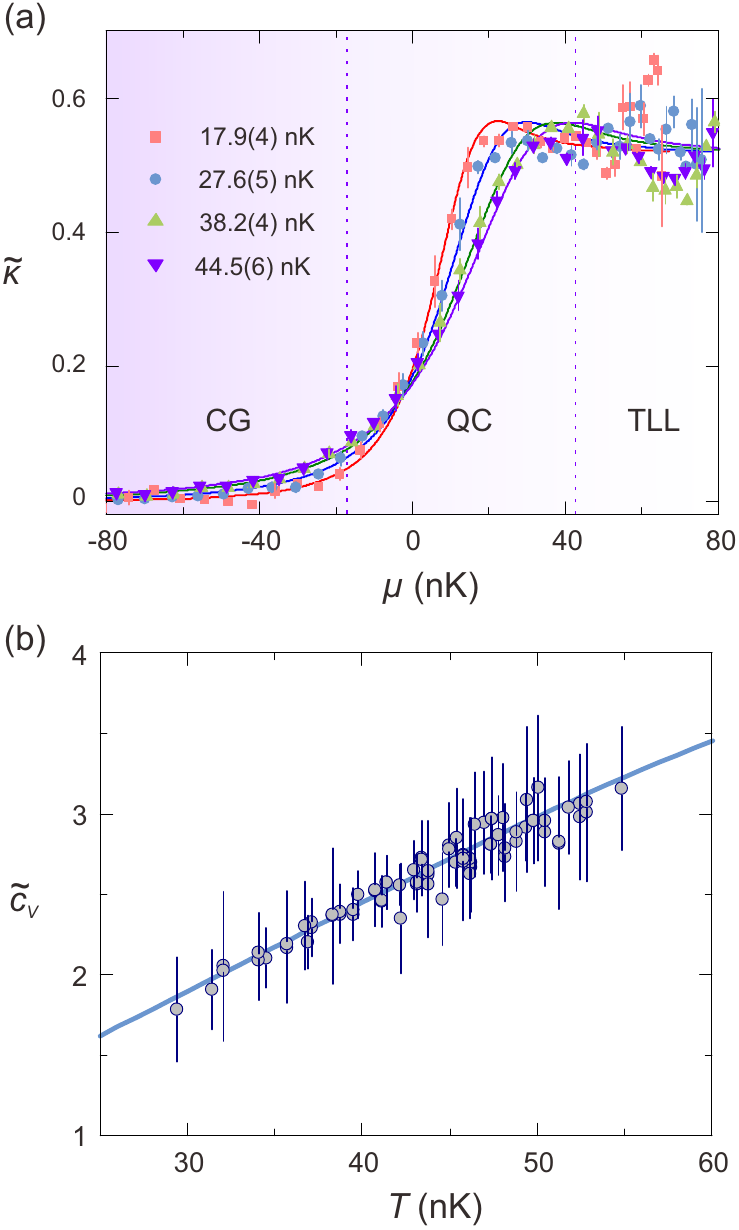}}\caption{Compressibility $\kappa$ and specific heat $c_V$. (a) The data are extracted from the density profiles, the solid curves are theoretical prediction based on the Yang-Yang equation. The dashed lines are boundaries of the quantum critical regime at $T = 44.5(6)$ nK. (b) The specific heat versus temperature at the chemical potential of $\mu = 60$ nK. The theoretical curve indicates the linear relation between $c_V$ and $T$. Here the error bars denote $\pm 1\sigma$ statistical errors.} \label{Fig.S4}
\end{figure}

Furthermore, we employ the compressibility $\kappa$ and the specific heat $c_V$ to describe the quantum critical behavior of the 1D system,
\begin{equation}
    \kappa=\left(\frac{\partial n}{\partial \mu}\right)_{\mu,c},\qquad c_{V} =T\left(\frac{\partial S}{\partial T}\right)_{\mu,c}. \label{thermodynamic}
\end{equation}
We apply similar discrete differential to get these two thermodynamics variables. The error bars of these derived quantities are deduced through uncertainty propagation. As shown in Fig. \ref{Fig.S4}(a), the compressibility get enhanced due to the quantum fluctuation in the quantum critical regime. While in the TLL regime, the compressibility at different temperatures almost approaches a constant. Meanwhile, the specific heat in Fig. \ref{Fig.S4}(b) follows a linear response to the temperature. Such temperature-independent compressibility and $T$-linear specific heat persist the collective behavior of the quantum liquid [34-36].

\section{Scaling functions and critical exponents}
In the vicinity of the quantum critical point $\mu_c = 0$, the 1D density obeys a universal scaling [5].
\begin{equation}
n(\mu,T) = T^{\frac{d}{z}+1-\frac{1}{\nu z}}{\cal{F}}\left(\frac{{\mu}-{\mu_c}}{T^{\frac{1}{\nu z}}}\right), \label{generic function}
\end{equation}
here the dimensionality is $d=1$. The ${\cal{F}}\left(x\right)$  is a generic function, where $z$ and $\nu$ are critical exponent and correlation length exponent. When $T \rightarrow 0$, the critical exponents given by the Yang-Yang equation are  $z=2$ and $\nu=0.5$.

We can define two scaled variables as $A(z,\nu,\mu,T) = n/ T^{1/z + 1- 1/\nu z}$ and $B(z,\nu,\mu,T) = \mu/T^{1/\nu z}$, then the scaling equation is simplified to $A(z,\nu,\mu,T) = {\cal{F}}[B(z,\nu,\mu,T)]$. The scaled curves at different temperatures should collapse to a single curve when the $z$ and $\nu$ are chosen correctly. We can define a function with respect to $z$ and $\nu$ to characterize the collapsing behavior,
\begin{equation}
 D(z, \nu) = \frac{1}{N M} \sum_{i=1}^{N} \sum_{j=1}^{M} \left[A (z, \nu, \mu_i ,T_j) - \overline{ A (z, \nu, \mu_i)} \right]^{2},
\end{equation}
where $\overline{A(z, \nu, \mu_i)}= \sum_{j=1}^M A(z, \nu, \mu_i, T_j)/M$ is the mean value. The minimum value of the function $D(z, \nu)$ corresponds to the best-fit critical exponents.

To evaluate the critical exponents, 6 curves (same as Fig.2) with temperatures ranging from $T = 17.9(4)$ nK to $T = 74.4(7)$ nK are taken into account. The errors of density measurements propagate to the function of $D(z, \nu)$, then map to the uncertainty of $z$ and $\nu$. With a 95$\%$ confident level of the uncertainty, the critical exponents are determined as $z= 2.3^{+0.6}_{-0.3}$, $\nu=0.56^{+0.07}_{-0.08}$. Our results agree with the critical exponents  $z=2$ and $\nu=1/2$ in the zero-temperature limit.

\begin{figure}[!b]
\centerline{\includegraphics[width=7cm]{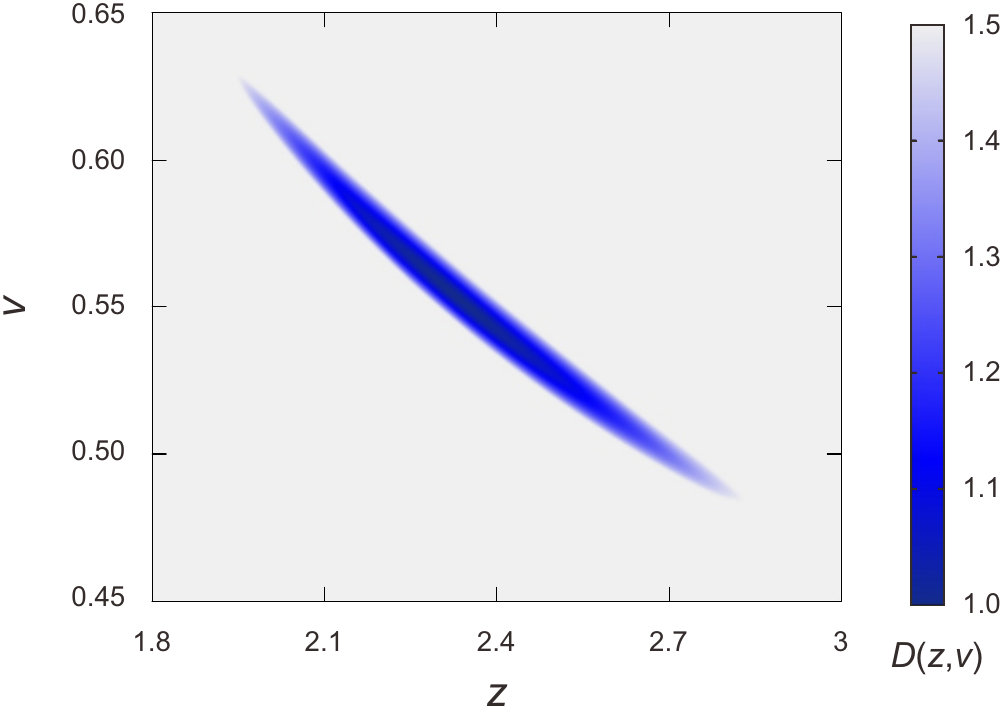}}\caption{Determination of the critical parameters. The color represents the value of $D(z, \nu)$. All the values of $D(z, \nu)$ are normalized to the minimum point of $D(z, \nu)$. The grey area is outside the 95$\%$ confidence level of the uncertainty.}\label{Fig.S5}
\end{figure}

Besides density, all the other thermodynamic quantities satisfy similar scaling laws in the QC regime. The full expressions of these scaling functions contain both singular and regular terms [2, 5], such as
\begin{equation}
\begin{split}
p(\mu,T) - p_r(\mu,T) = T^{\frac{d}{z}+1}{\cal{G}}\left(\frac{{\mu}-{\mu_c}}{T^{\frac{1}{\nu z}}}\right), \\
S(\mu,T) - S_r(\mu,T) = T^{\frac{d}{z}}{\cal{H}}\left(\frac{{\mu}-{\mu_c}}{T^{\frac{1}{\nu z}}}\right). \label{generic function of p and S}
\end{split}
\end{equation}
Here, $\cal{G}$ and $\cal{H}$ are scaling functions for pressure $p$ and entropy density $S$, respectively. For the critical behavior of the vacuum-to-TLL transition, the regular part of density is negligible in the low-temperature limit. While the regular terms for pressure or entropy are weakly dependent on temperature, almost independent on the chemical potential. The universal scaling of pressure and entropy are shown in Fig.2, where the regular parts are calculated by the shifts at the critical point $p_r(\mu_c,T)$, $S_r(\mu_c,T)$. The singular part of the thermodynamic variable determines the shape of the scaling function, while the regular part compensates the bias between curves of different temperatures.

\section{Measuring the sound velocity}

\begin{figure}[!htb]
\centerline{\includegraphics[width=6.5cm]{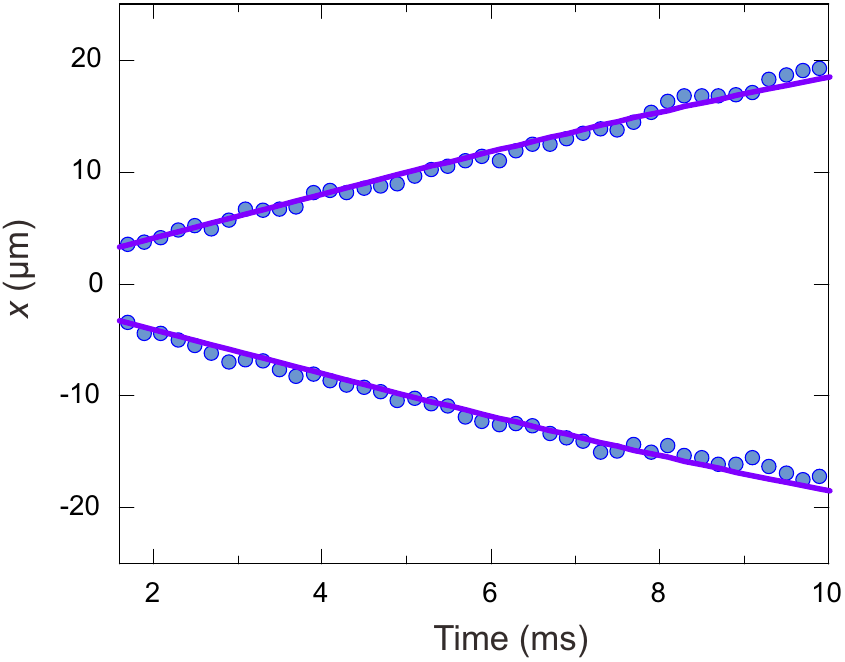}} \caption{Sound propagation in the 1D gases. The initial perturbations are in Gaussian shape with $\eta = 0.17(1)$ and  $w = 4.0(4)\ \mu m$. The solid curves are the theoretical fittings with a single variable $v_s(t_0)$ based on the relation $x(t) = R_{_{TF}} \sin[v_s(t_0)t/R_{_{TF}}]$. } \label{Fig.S6}
\end{figure}

The 1D axial trap is along the same direction as the gravity, which is compensated by a magnetic gradient field. Such magnetic gradient produces a spatial-dependent Zeeman splitting, making the MW transition frequency also spatial-dependent. When the MW frequency resonates with the center of the cloud, the central atoms can be transfer from $\Ket{F=1,m_F=-1}$ to $\Ket{F=2,m_F=-2,-1,0}$. Followed by a resonant light pulse (10 $\mu$s duration) to remove the atoms in $\Ket{F=2}$, density dips are created in the center of the 1D tubes. The shape of these defects is approximately Gaussian $\eta n e^{-x^2/2 w^2}$, where the Gaussian width $w$ and the relative amplitude $\eta$ are tailored by adjusting the MW strength and the transition routes. In general, a $\pi$ transition (to $\Ket{F=2,m_F=-1}$ state) pulse in 30 $\mu s$ duration generates a dip with $\eta = 0.17(1)$ and $w = 4.0(4)\ \mu m$ (Fig. \ref{Fig.S6}). The dip creation time can be neglected comparing with the propagation of sound. The healing length is approximately 0.1 $\mu$m at the peak density of 1D tubes, much smaller than the perturbation
width [27].

The negative perturbation splits into two parts and symmetrically propagate along the 1D tube with the speed of sound. Since the sound velocity is density-dependent as $v_s \propto \sqrt{n}$, the propagating distance obeys the relation of  $x(t) = R_{_{TF}} \sin[v_s(t_0)t/R_{_{TF}}]$ with $R_{_{TF}}$ the Thomas-Fermi radius [28]. However, the density-dependent relation also indicates that the center of the dip moves slower than the edges. A shock wave will form when the trailing edge overrun the center at a certain time [29]. The time to form a shock wave is approximately $t_s \sim 8 w/v_s\eta$, which is 86 ms for the above 30$\mu s$ pulse setting. Even before forming of the shock wave, the shape of the sound wave has some distortion. In the experiment, the details and the shape of the sound wave are smoothed due to the noise and averaging. The mean data of the propagating wave are acquired by averaging of 20 images in the same experimental setting. We use a double-peak Gaussian function to fit the mean data and find the locations of the dips. Fig. \ref{Fig.S6} shows the typical sound propagation behavior.

Measuring the sound velocity with an infinitesimally small perturbation is limited by the finite signal-to-noise ratio. However, we can extrapolate $v_s(\eta = 0)$ from the sound velocities with finite perturbations based on the relation of $v_s(\eta) = v_s(0)\sqrt{1-\eta/2}$. As shown in Fig.4 of the main text, $v_s$ in different excitation ratios $\eta$ consists with this relation. Here, the propagation time is limited to be less than 4 ms to ensure the propagating is inside the TLL regime. The up-level is chosen as $\ket{F=2,m_F=0}$ to get the largest perturbation width $w$. For the deepest excitation $\eta=0.74$ and $w=2.5 \ \mu$m, the shock wave forming time $t_s = 12$ ms is longer than the duration of the sound propagation.

\section{Measuring the momentum distribution}
\begin{figure}[!htb]
\centerline{\includegraphics[width=6.9cm]{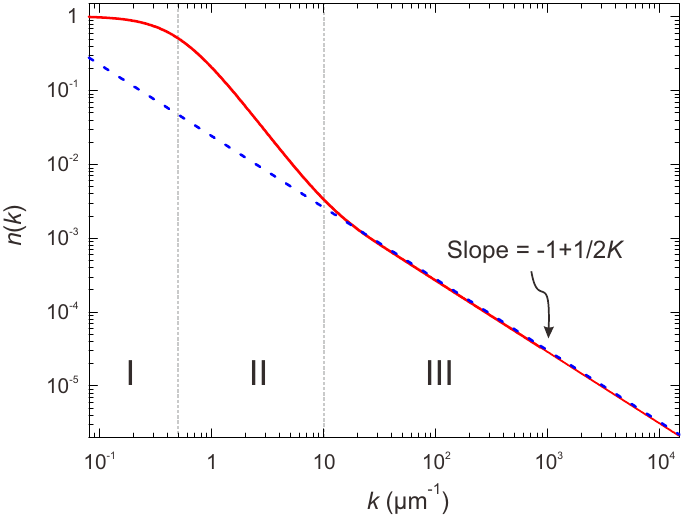}}\caption{Theoretical momentum distribution of a degenerate gas. The temperature of this 1D gas is 40 nK and the Luttinger parameter is $K =15.9$. Here red curve is the theoretical prediction in this condition. While the blue dashed curve with a linear slope $-1+1/2K$ is a guide to the eye. The experimental accessible regime are I and II. While in the regime III, the momentum distribution has an asymptotic power-law decay that only depends on the Luttinger parameter $K$ [23].} \label{Fig.S7}
\end{figure}

For a 1D uniform system, the momentum spectrum $n(k)$ along the weakly confined direction is the Fourier transform of the first-order correlation function. As demonstrated in [23], the momentum distribution has a form like,
\begin{equation}
  n(k)\simeq A(K) \text{Re}\left[\frac{\Gamma(1/4K + \text{i}kl_{\phi}/2K)}{\Gamma(1-1/4K+\text{i}kl_{\phi}/2K)}\right],
\end{equation}
here $A(K)$ is a $K$-dependent parameter. In the harmonic potential, the momentum distribution of a quantum gas can be deduced by introducing an effective Luttinger parameter. Since the correlation length of our experiment is much smaller than the system size, LDA can be employed to describe the 1D system. At a temperature of 40(1) nK and $\bar{K} = 15.9$, the theoretical prediction of the momentum distribution is shown in a log-log plot as Fig. \ref{Fig.S7}. Such a curve has three different regions, one is the momenta below $k < 1/l_{\phi}$, another is the intermediate region $1/l_{\phi} \leq k \leq 20/l_{\phi}$ and the last is the region of high momenta $k > 40/l_{\phi}$. At $k < 1/l_{\phi}$, thermal fluctuations disturb the long-range correlations and dominate this part. In the intermediate region, the TLL shows its effect and an algebraic decay emerges in the distribution. At even higher momenta, a power-law decay with a slope of $-1 + 1/2K$ prevails, which is a characteristic behavior of the TLL.

In the experiment, the momentum distribution can be measured after the time-of-flight mapping. Limited by the system size and the signal-to-noise ratio, infinite long-expand is not allowed in quantum gas experiments. However, the initial size of the 1D gas would affect the mapping if the expansion time is not long enough. To solve this confliction, we utilize a focusing technique during the time-of-flight. In order to avoid the effect of the gravity during the expansion, we build an optical trapping potential with the same $\omega_{\perp}$ as described in Fig.1(a) but switch the red-lattice beams to the $x$ direction, thereby create an array of 1D tubes with the weak confinement along $y$. After preparing the atoms in a thermal equilibrium state, we suddenly release the optical confinements and let the cloud evolve in a weak magnetic potential. The magnetic trap has a harmonic trapping frequency $\omega_h = 2\pi \times 10.0(2)$ Hz along the $y$ direction. The sample is allowed to expand for a quarter period and then the initial momentum distribution is mapped to a spatial density distribution with $k = m \omega_y y/2\pi$. In addition, we probe this 1D gas using \emph{in situ} imaging and obtain a temperature of $40(1)$ nK. Since the accessible range is limited by the healing length, the power-law decay of high momenta cannot be resolved here. However, the agreement between the experiment and the theoretical prediction indicates that the finite-temperature TLL dominates the behavior of this 1D Bose gas.

\section{Luttinger parameter and Wilson ratio}

The low energy physics of the 1D Bose gases can be described by an effective Hamiltonian.
\begin{equation}
  H= \int d x \left[ \frac{\pi v_{s} K}{2} \Pi^{2} + \frac{v_{s}}{2 \pi K} (\partial_x \phi )^{2} \right],
\end{equation}
where the canonical momenta $\Pi$  and the phase $\phi$ obey the standard Bose commutation relation. Here the Luttinger parameter $K$ and sound velocity $v_{s}$ characterize the low-energy 1D system and determine the long-distance asymptotic behaviour of correlation functions. For a homogenous 1D Bose gas with length $L$ and number of particles $N$, the Luttinger parameter is given by $K= v_{s}/v_{N}$. Here the sound velocity is $v_{s} = \sqrt{\left(\partial^{2} E/\partial L^{2}\right)L^{2}/Nm }$, and stiffness is $v_{N} = \left(\partial^{2} E/\partial N^{2}\right) L/\pi \hbar$. Since the ground state energy $E$ is related to the energy density function $e( \gamma )$ as
\begin{equation}
  E= \frac{\hbar^{2}}{2m} \frac{N^{3}}{L^{2}} e ( \gamma ),
\end{equation}
then the sound velocity and stiffness are
\begin{equation}
\begin{split}
v_s &=\frac{\hbar n}{m}\sqrt{3e-2\gamma \frac{d e}{d \gamma}+\frac{1}{2}\gamma^2 \frac{d^2 e}{d^2 \gamma}},\\
v_N &=\frac{\hbar n}{\pi m}(3e-2\gamma \frac{d e}{d \gamma}+\frac{1}{2}\gamma^2 \frac{d^2 e}{d^2 \gamma}).
\end{split}
\end{equation}
We can get a simple relation between the Luttinger parameter and the sound velocity,
\begin{equation}
K=\frac{v_s}{v_N}=\frac{\pi}{\sqrt{3e-2\gamma \frac{d e}{d \gamma}+\frac{1}{2}\gamma^2 \frac{d^2 e}{d^2 \gamma}}}=\frac{\pi\hbar n}{m v_s}
\end{equation}
By assuming an effective local chemical potential in the non-uniform system (based on LDA), we can extend such a relation to the 1D harmonic trapped Bose gas. Therefore, an effective $K$ can be deduced experimentally by measuring the mean density and the averaged sound velocity.

\begin{figure}[!htb]
\centerline{\includegraphics[width=6.5cm]{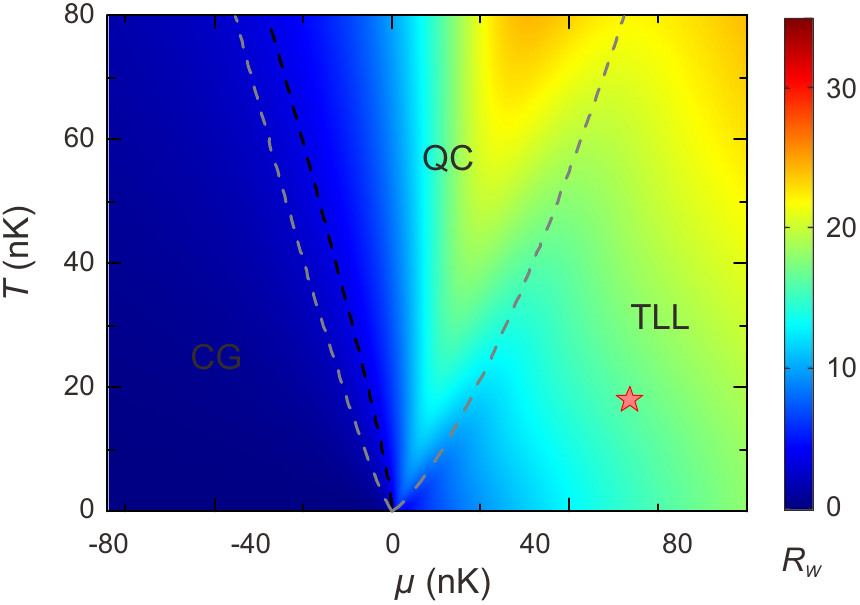}}\caption{Phase diagram of $R_W$ in the $\mu-T$ plane. The dashed grey lines are the peak values of specific heat, which separate different regimes. The dashed black curve represents the quantum degenerate boundary, at which the thermal de Broglie wavelength equals the average atomic spacing. The red star locates the lowest temperature of our experiments.  } \label{Fig.S8}
\end{figure}

However, the Luttinger parameter $K$ cannot describe the 1D properties other than the TLL regime. In the quantum critical regime, the linear dispersion is no longer valid due to the existence of quantum fluctuations. Adopting another dimensionless parameter from strongly correlated Fermi liquid $--$ the Wilson ratio [37], we can properly characterize different 1D regimes. In Fermi liquid theory, Wilson ratio is a ratio defined by the magnetic susceptibility and the specific heat. Since the compressibility $\kappa$ in the 1D Bose gas plays almost the same role as the susceptibility in Fermi liquid, we can define a Wilson-like ratio [34, 35, 37].
\begin{equation}
  R_{W} = \frac{\pi^{2} k_{B}^{2}}{3} \frac{\kappa}{c_{V} /T}.
\end{equation}
The TLL phase is a Galilean invariant system, in which the specific heat $c_{V}$ relates to temperature and sound velocity as,
\begin{equation}
  c_{V} = \frac{\pi k_{B}^{2} }{3 \hbar} \frac{T}{v_{s}} \label{spc}.
\end{equation}
Together with the relation $\kappa = 1/(\hbar \pi v_{N})$, we obtain a remarkable correlation between the Wilson ratio and the Luttinger parameter in the 1D homogenous gas,
\begin{equation}
R_{W} =K.
\end{equation}
This builds up an intriguing connection between the low-energy physics of the Fermi liquid and the Luttinger liquid theory [34, 35]. In the Fig. \ref{Fig.S8}, the averaged Luttinger parameters $\bar{K}$ at the mean chemical potentials of $63.5$ nK and $65.3$ nK are 16.9 and 17.2, respectively. The derived Wilson ratios approach the averaged Luttinger parameters in the TLL regime. The Wilson ratio can feature the 1D Bose gases at different regimes, ranging from the Luttinger liquid to the classical gas. Such a relation also provides a measurable method for the dimensionless parameters in all the 1D many-body systems. As shown in Fig. \ref{Fig.S8}, the phase diagram of Wilson ratio can map out the quantum critical properties. The ``V" shape of the Wilson ratio also indicate three fluctuation regimes. The right crossover branch matches the valley of Wilson ratio and give the temperature scale $T^{*}\sim |\mu-\mu_c|^{z\nu}$. The left crossover temperatures $T^{*}$ can be given by the thermal wavlength $\lambda_T \sim 1/n$. The crossover boundaries given by the specific heat and the Wilson ratio are consistent with each other.

\end{document}